\documentclass[letterpaper]{article} 
\usepackage{aaai2027}  
\usepackage[hyphens]{url}  
\usepackage{graphicx} 
\usepackage{natbib}  
\usepackage{caption} 
\usepackage{amsmath}
\usepackage{amssymb}
\usepackage{amsfonts}

\usepackage[table]{xcolor}
\usepackage{booktabs}
\usepackage{multirow}
\usepackage{array}
\usepackage{tabularx}
\usepackage{makecell}
\usepackage{arydshln}

\usepackage{algorithm}
\usepackage{algpseudocode}

\usepackage{newfloat}
\usepackage{listings}
\DeclareCaptionStyle{ruled}{labelfont=normalfont,labelsep=colon,strut=off} 
\floatstyle{ruled}
\newfloat{listing}{tb}{lst}{}
\floatname{listing}{Listing}

\usepackage{booktabs}

\title{Disentangled Contrastive Learning for Zero-Shot Multilingual Dense Retrieval}
\author{
   Chao Huang\textsuperscript{\rm 1,\rm 2},
   Yufeng Chen\textsuperscript{\rm 1,\rm 2},
   Changhao Guan\textsuperscript{\rm 1,\rm 2},
   Guang Yang\textsuperscript{\rm 1,\rm 2},
   Dongze Chen\textsuperscript{\rm 2},
   Kaiyu Huang\textsuperscript{\rm 1,\rm 2}\corresponding
}
\affiliations{
    
    \textsuperscript{\rm 1}Key Laboratory of Big Data \& Artificial Intelligence in Transportation \\
    (Beijing Jiaotong University), Ministry of Education \\
    \textsuperscript{\rm 2}School of Computer Science and Technology, Beijing Jiaotong University \\
    huangchao@bjtu.edu.cn, kyhuang@bjtu.edu.cn
}

\begin{document}

\nocopyright
\maketitle

\begin{abstract}
Multilingual dense retrieval aims to handle queries and documents across different languages based on a unified retriever model. 
The challenge lies in enabling robust retrieval transfer to low-resource languages where annotated retrieval data is often scarce.
Although previous studies transfer high-resource supervision to low-resource languages in multilingual semantic representation learning, the shared representation often entangles semantic and linguistic features, which may interfere with optimizing semantic relevance for retrieval.
Different from existing methods that focus on learning language-agnostic semantic features under such entanglement, we propose a disentangled contrastive learning~(DCL) method for multilingual dense retrieval by separating multilingual representations into semantic and linguistic subspaces. Specifically, we design disentangled optimization objectives based on hierarchical semantic alignment and language debiasing contrastive learning. By aligning retrieval-relevant semantics across languages at both sentence and token levels while capturing language-specific variations in the linguistic subspace, these objectives reduce language-induced interference in semantic matching. We jointly optimize them with the retrieval objective to facilitate stable zero-shot transfer from English supervision to multilingual dense retrieval.
Extensive experiments on mMARCO and MIRACL show that our method consistently outperforms several strong baselines, demonstrating its effectiveness and generalization ability.
\end{abstract}

\section{Introduction}

Multilingual dense retrieval~\cite{nie2010cross,zhang2023toward} 
aims to learn a retriever in a shared semantic space so that the model can handle queries and documents in different languages.
However, constructing a multilingual dense retriever is not trivial due to the lack of annotated data, especially for low-resource languages.
Thus, 
a common practice is to train a multilingual pre-trained model on high-resource language annotated data (e.g., English) and then generalize the retrieval ability to low-resource languages~\cite{macavaney2020teaching,lin2023maggretriever,litschko2023boosting}. 

Although this paradigm yields some gains, its performance remains limited by two main aspects. First, multilingual pre-trained models suffer from inadequate cross-lingual semantic alignment. Due to the variations of linguistic features (e.g., grammatical structure, vocabulary and word order), sentences in different languages with equivalent semantics may be mapped to different regions in the shared representation space~\cite{schuster2019cross,tanwar2023multilingual,hammerl2024understanding}, which makes it difficult for retrieval supervision learned from high-resource languages to effectively generalize to low-resource languages. Second, multilingual representations often entangle retrieval-relevant semantic features with linguistic features. When fine-tuning a retriever with high-resource language data, models may over-rely on language-relevant patterns rather than learning purely transferable semantic relevance, further weakening zero-shot transfer capabilities across languages for retrieval~\cite{huang2021disentangling,jian2022non}.

Previous studies have predominantly concentrated on learning the effective shared representation space to enhance transferability~\cite{wang2022english,chen2024bge}, including modular vector representations~\cite{zhang2023toward,louis2025colbert} and synthesizing data~\cite{huang2023soft,thakur2024leveraging}.
However, an excessive emphasis on cross-lingual semantic alignment may exacerbate the entanglement between retrieval-relevant semantic features and language-specific features.
Disentangling confounded semantics while optimizing the shared cross-lingual semantic representation space is crucial for improving transfer performance in multilingual retrieval.

\begin{figure*}
    \centering
    \includegraphics[width=1\linewidth]{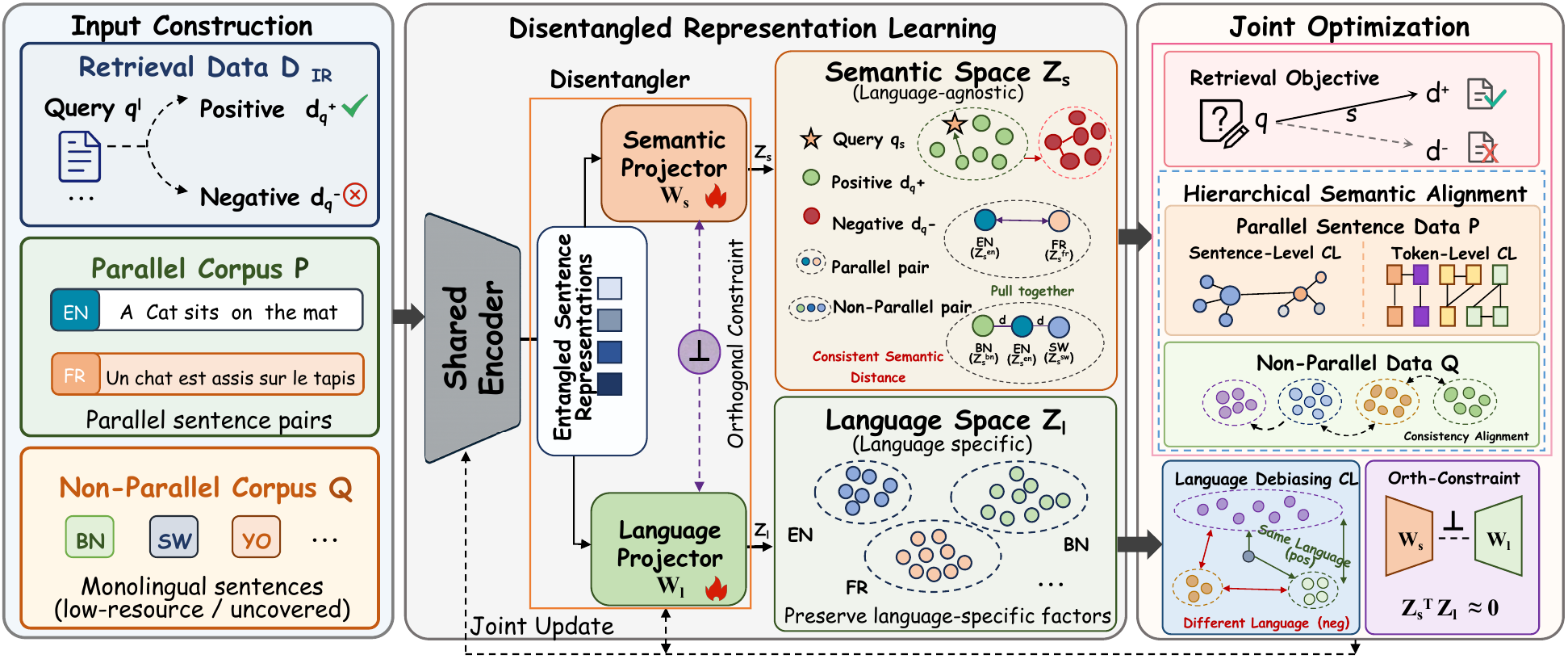}
    \caption{Overview of our disentangled framework, including (i) a language-semantic disentangler separating semantic and linguistic features, (ii) hierarchical semantic alignment performing multi-aspect semantic alignment, and (iii) language debiasing contrastive learning to capture language-specific variations in the linguistic subspace.}
    \label{fig:placeholder}
\end{figure*}

The entanglement of semantic and linguistic features can be formulated as a feature separation problem. Motivated by this, a potential solution is to design disentanglement modules and incorporate them into multilingual retrievers.
Although disentangled representation learning has demonstrated its ability to decompose entangled sentence representations and exploit the disentangled semantic features to enhance natural language understanding tasks~\cite{chen2019multi,huang2021disentangling,wu2022learning,zhang2024learning,wang2022causal,wang2024semantics}, extending it to multilingual dense retrieval is non-trivial.
In multilingual dense retrieval, the disentangled semantic subspace must preserve fine-grained query-document relevance signals that are essential for ranking while maintaining cross-lingual alignment, and the linguistic subspace should absorb those language-specific differences that interfere with semantic matching. Thus, directly using existing disentanglement objectives may lead to unstable performance of zero-shot multilingual dense retrieval on low-resource languages or out-of-domain datasets.

To achieve a retrieval-oriented disentanglement objective, we propose a disentangled contrastive learning~(DCL) framework to jointly promote cross-lingual semantic alignment and semantic-linguistic feature separation for multilingual dense retrieval.
In the two disentangled spaces, the method learns multi-granular retrieval-relevant representations in the semantic space and language-specific representations in the linguistic space. The framework consists of three stages.
 The first stage is to disentangle the sentence representation into semantic and linguistic features through two projection matrices. Then, we design the disentangled optimization objectives, including hierarchical semantic alignment and language debiasing contrastive learning. 
The former uses sentence-level hard negatives to preserve retrieval distinctions, token-level alignment to capture fine-grained cross-lingual correspondences, and consistency constraints to transfer the aligned structure to languages without parallel data. In this way, the semantic subspace learns language-invariant yet ranking-sensitive semantic structures that support fine-grained cross-lingual relevance matching. The latter directs language-specific patterns to the linguistic subspace by pulling same-language representations together and pushing different-language ones apart. Together with the orthogonal constraint, they reduce language leakage and focus the semantic subspace on language-invariant relevance signals.
Finally, we jointly optimize the multilingual dense retrieval and disentanglement objectives to facilitate stable zero-shot transfer from English supervision to multilingual dense retrieval. Experimental results on the multilingual retrieval datasets mMARCO and MIRACL demonstrate the effectiveness of our method. 
Further analysis quantifies the contribution of each key component and provides representation visualizations to examine whether the learned semantic and linguistic subspaces exhibit the expected language-invariant and language-aware properties.

Our contributions are summarized as follows: 
1) We propose a disentangled contrastive learning~(DCL) framework for multilingual dense retrieval, which successfully achieves retrieval-oriented semantic-linguistic disentanglement.
2) We design two complementary disentangled optimization components: hierarchical semantic alignment promotes sentence-level and token-level cross-lingual alignment of retrieval-relevant semantic features, while language debiasing contrastive learning encourages the linguistic subspace to capture language-specific variations.  
3) Experimental results show that our approach outperforms strong baselines in terms of effectiveness and generalization for multilingual retrieval in-domain and out-of-domain, without requiring any multilingual retrieval data, obtaining consistent gains across resource levels and language families.

\section{Related Work}

\subsection{Zero-shot Multilingual Information Retrieval}
Zero-shot multilingual information retrieval aims to transfer the retrieval capability learned from English supervision to low-resource languages~\cite{litschko2023boosting,lin2023maggretriever}. To achieve this goal, a series of methods have been proposed to improve the zero-shot cross-lingual retrieval ability of multilingual retrievers, which can be roughly divided into two categories. First, some studies develop embedding models to fine-tune multilingual retrievers~\cite{devlin2019bert,conneau2019cross,chi2021infoxlm,feng2022language,wang2022english}. They aim to leverage multilingual pre-trained language models to enable zero-shot transfer in multilingual dense retrieval. Second, other studies introduce language-specific modules into multilingual retrievers to improve cross-lingual transfer by adapting the retriever to different languages~\cite{zhang2023toward,louis2025colbert}.
Different from existing studies that focus on learning language-agnostic semantic features under the entanglement of semantic and linguistic information, our method explores how to use the separated semantic features to facilitate zero-shot cross-lingual transfer from the perspective of disentangled optimization.
\subsection{Disentangling for Cross-lingual Alignment}
Cross-lingual semantic alignment~\cite{cao2020multilingual,ouyang2021ernie} is one of the core goals of multilingual sentence representation learning, which makes sentences with similar semantics in different languages closer to each other in a shared vector space. Although multilingual pre-trained models have achieved strong cross-lingual capabilities during the pre-training or further fine-tuning stage, the entangled semantic and linguistic features may impair the model performance. Recent studies~\cite{huang2021disentangling,wu2022learning,zhang2024learning} attempt to perform cross-lingual semantic alignment from the perspective of disentanglement, such as using autoencoders or SVD to extract linguistic features, and leveraging language-agnostic semantic information for task-specific training. However, their disentanglement objectives are relatively simple and mostly applied to natural language understanding and generation tasks. Different from them, we aim to design a disentangled optimization objective and adapt it to multilingual dense retrieval.

\section{Methodology}
As shown in Figure~\ref{fig:placeholder}, we propose a disentangled contrastive learning framework with hierarchical
semantic and linguistic alignment to facilitate effective zero-shot cross-lingual transfer in multilingual dense retrieval, which consists of three main components: i) a language-semantic disentangler for separating the entangled representation into semantic and language-specific subspaces, ii) hierarchical semantic alignment by performing cross-lingual alignment at both sentence and token levels with parallel and non-parallel corpora and iii) capturing language-specific variations in the linguistic subspace by language debiasing contrastive learning.

\subsection{Language-semantic Disentangler}

For multilingual dense retrieval, the encoder representation usually entangles task-related semantic information and language-specific features, such as grammatical patterns and word-order variations. Such entanglement may introduce language-dependent bias into the shared retrieval space and hinder cross-lingual semantic alignment. To alleviate this issue, we introduce a language-semantic disentangler, which projects the original sentence representation $\mathbf{z}$ into a semantic component $\mathbf{z}_s$ and a language-specific component
$\mathbf{z}_l$:
\begin{equation}
\mathbf{z}_s = W_s \mathbf{z}, 
\quad 
\mathbf{z}_l = W_l \mathbf{z}
\end{equation}
where $W_s$ and $W_l$ are trainable parameters of the semantic and language projection heads, respectively. The semantic component $\mathbf{z}_s$ is used to capture retrieval-relevant semantic information, while $\mathbf{z}_l$ preserves language-specific features. To reduce the correlation between semantic and language-specific subspaces, we impose an orthogonal constraint on the batch-level projected representations. Given a mini-batch, we stack the semantic and language-specific representations as $\mathbf{Z}_s,\mathbf{Z}_l\in\mathbb{R}^{B\times d'}$, where $B$ is the batch size. The orthogonal loss is defined as
\begin{equation}
\mathcal{L}_{\mathrm{orth}}
=
\left\|
\frac{1}{B}\mathbf{Z}_s^{\top}\mathbf{Z}_l
\right\|_{F}^{2}
\end{equation}
where $\|\cdot\|_F$ denotes the Frobenius norm.

\subsection{Hierarchical Semantic Alignment}

In practice, parallel corpora are often unavailable for many low-resource languages. To enable cross-lingual transfer under limited supervision, we use the available parallel corpus $\mathbb{P}$ for direct semantic alignment and the non-parallel corpus $\mathbb{Q}$ for indirect consistency constraints. Specifically, we perform hierarchical semantic alignment on the language-agnostic component $\mathbf{z}_s$ at both sentence and token levels. For parallel data, we align semantically equivalent sentence pairs and their aligned tokens, yielding $\mathcal{L}_{\mathrm{sentCL}}^{\mathbb{P}}$ and $\mathcal{L}_{\mathrm{tokenCL}}^{\mathbb{P}}$. For non-parallel data, we encourage monolingual representations to maintain consistent semantic distances to parallel sentence pairs, yielding $\mathcal{L}_{\mathrm{sentCL}}^{\mathbb{Q}}$ and $\mathcal{L}_{\mathrm{tokenCL}}^{\mathbb{Q}}$. We further mine in-batch hard negatives for parallel sentence-level alignment to improve fine-grained semantic discrimination.

\noindent \textbf{Sentence-level Cross-lingual Alignment}. 
For semantically equivalent sentences in different languages, we aim to learn a consistent semantic representation. Given a parallel sentence pair $(x_i, x_j)$, we obtain their semantic representations through the encoder $f(\cdot)$ and the semantic projection layer as
\begin{equation}
\mathbf{s}_i = W_s f(x_i), 
\quad 
\mathbf{s}_j = W_s f(x_j)
\end{equation}

To provide more challenging negatives, we mine the top-$K$ most similar
samples in the batch, excluding the current parallel pair, as their hard negative sets, denoted as $\mathcal{N}_i=\mathrm{TopK}(\{\mathbf{s}_k\mid k\neq i,j\},
\mathrm{sim}(\mathbf{s}_i,\mathbf{s}_k),K)$ and
$\mathcal{N}_j=\mathrm{TopK}(\{\mathbf{s}_k\mid k\neq i,j\},
\mathrm{sim}(\mathbf{s}_j,\mathbf{s}_k),K)$, respectively.

For compactness, we first define
$\delta(\mathbf{u},\mathbf{v})=\exp(\mathrm{sim}(\mathbf{u},\mathbf{v})/\tau)$
as the temperature-scaled exponential similarity between two representation
vectors $\mathbf{u}$ and $\mathbf{v}$. Based on this, the contrastive loss for
an anchor $\mathbf{x}$, its positive sample $\mathbf{x}^{+}$, and a negative set
$\mathcal{N}$ is defined as
\begin{equation}
\resizebox{0.97\linewidth}{!}{$
\begin{aligned}
\ell(\mathbf{x},\mathbf{x}^{+},\mathcal{N})
=
-\log
\frac{
\delta(\mathbf{x},\mathbf{x}^{+})
}{
\delta(\mathbf{x},\mathbf{x}^{+})
+
\sum_{\mathbf{x}^{-}\in\mathcal{N}}
\delta(\mathbf{x},\mathbf{x}^{-})
}
\end{aligned}
$}
\end{equation}

Based on the selected hard negative samples, for $N$ pairs of parallel sentences
within a batch, we construct the sentence-level contrastive loss
$\mathcal{L}_{\mathrm{sentCL}}^{\mathbb{P}}$ as
\begin{equation}
\resizebox{0.97\linewidth}{!}{$
\begin{aligned}
\mathcal{L}_{sentCL}^{\mathbb{P}}
=
\frac{1}{2N}
\sum_{(i,j)\in\mathbb{P}}
\left[
\ell(\mathbf{s}_i,\mathbf{s}_j,\mathcal{N}_i)
+
\ell(\mathbf{s}_j,\mathbf{s}_i,\mathcal{N}_j)
\right]
\end{aligned}
$}
\end{equation}

\noindent \textbf{Token-level Cross-lingual Alignment.} 
Sentence-level cross-lingual alignment alone may be insufficient to capture fine-grained semantic correspondence, especially for low-resource languages with substantial variations in word order and morphology. To address this issue, we further introduce token-level cross-lingual alignment. Given a parallel sentence pair $(x_i,x_j)$, we denote their tokenized forms as $x_i=\{u_1,u_2,\cdots,u_m\}$ and $x_j=\{v_1,v_2,\cdots,v_n\}$. Then, we use awesome-align~\cite{dou2021word} to obtain token-level alignment indices, denoted as $\mathcal{T}_{ij}=\{(a,b)\mid u_a\leftrightarrow v_b,\ 1\leq a\leq m,\ 1\leq b\leq n\}$.

For all parallel sentence pairs in the batch, we merge their alignment indices as $\mathcal{T}=\bigcup_{(x_i,x_j)\in\mathcal{B}_{\mathbb{P}}}\mathcal{T}_{ij}$, with $|\mathcal{T}|=T$. For each aligned index pair $(a_t,b_t)\in\mathcal{T}$, we take the corresponding contextualized token representations from the encoder outputs, denoted as $\mathbf{w}_{u_t}$ and $\mathbf{w}_{v_t}$.

We denote all aligned token representations in the batch as
$\mathcal{W}^{\mathbb{P}}=\{\mathbf{w}_{u_t},\mathbf{w}_{v_t}\}_{t=1}^{T}$,
and define the token-level negative set as
$\mathcal{N}^{w}_{t}=\mathcal{W}^{\mathbb{P}}\setminus
\{\mathbf{w}_{u_t},\mathbf{w}_{v_t}\}$. The token-level contrastive loss is
defined as
\begin{equation}
\begin{aligned}
\mathcal{L}_{tokenCL}^{\mathbb{P}}
=
\frac{1}{2T}
\sum_{t=1}^{T}
\Big[
&\ell(\mathbf{w}_{u_t},\mathbf{w}_{v_t},\mathcal{N}^{w}_{t})
\\
&+
\ell(\mathbf{w}_{v_t},\mathbf{w}_{u_t},\mathcal{N}^{w}_{t})
\Big]
\end{aligned}
\end{equation}
Finally, the hierarchical semantic contrastive loss for parallel data is
defined as
\begin{equation}
\mathcal{L}_{\mathrm{HsemaCL}}^{\mathbb{P}}
=
\mathcal{L}_{\mathrm{sentCL}}^{\mathbb{P}}
+
\lambda_t \mathcal{L}_{\mathrm{tokenCL}}^{\mathbb{P}}
\end{equation}
where $\lambda_t$ is the weight hyperparameter for the token-level loss.

Similarly, for low-resource languages without parallel corpora, we construct indirect semantic consistency constraints based on the available parallel sentence pairs. Given a monolingual sentence $x_k\in\mathbb{Q}$, we obtain its semantic representation $\mathbf{s}_k$ through the encoder and the semantic projection layer. Since $x_k$ has no explicit parallel counterpart, we encourage it to have comparable semantic distances to the two sides of a parallel sentence pair $(x_i,x_j)$, i.e.,
$\mathrm{sim}(\mathbf{s}_i,\mathbf{s}_k)\approx
\mathrm{sim}(\mathbf{s}_j,\mathbf{s}_k)$.

For compactness, we define
$\rho(\mathbf{x};\mathbf{a},\mathbf{b})
=
\delta(\mathbf{x},\mathbf{a})/
(\delta(\mathbf{x},\mathbf{a})+\delta(\mathbf{x},\mathbf{b}))$
as the normalized pairwise similarity between an anchor $\mathbf{x}$ and two reference representations $\mathbf{a}$ and $\mathbf{b}$. Within the same batch, we control the number of parallel sentence pairs and low-resource monolingual sentences to be $N$, and construct the sentence-level consistency loss $\mathcal{L}_{sentCL}^{\mathbb{Q}}$
as
\begin{equation}
\begin{aligned}
\mathcal{L}_{sentCL}^{\mathbb{Q}}
=
-\frac{1}{N^2}
\sum_{(i,j)\in\mathbb{P}}
\sum_{k\in\mathbb{Q}}
\Big[
&\log \rho(\mathbf{s}_k;\mathbf{s}_i,\mathbf{s}_j)
\\
&+
\log \rho(\mathbf{s}_k;\mathbf{s}_j,\mathbf{s}_i)
\Big]
\end{aligned}
\end{equation}

At the token level, we apply the same consistency constraint to the token
representations of non-parallel sentences. Let
$\mathcal{W}^{\mathbb{Q}}=\{\mathbf{w}_{q_k}\}_{k=1}^{K}$ denote the token
representations from non-parallel data, and
$(\mathbf{w}_{u_t},\mathbf{w}_{v_t})\in\mathcal{W}^{\mathbb{P}}$ denote an
aligned token pair from parallel data. The token-level consistency loss is
$\mathcal{L}_{tokenCL}^{\mathbb{Q}}$ defined as
\begin{equation}
\begin{aligned}
\mathcal{L}_{tokenCL}^{\mathbb{Q}}
=
-\frac{1}{TK}
\sum_{t=1}^{T}
\sum_{k=1}^{K}
\Big[
&\log \rho(\mathbf{w}_{q_k};\mathbf{w}_{u_t},\mathbf{w}_{v_t})
\\
&\hspace{-1.5em}+
\log \rho(\mathbf{w}_{q_k};\mathbf{w}_{v_t},\mathbf{w}_{u_t})
\Big]
\end{aligned}
\end{equation}

Finally, the hierarchical semantic contrastive loss for non-parallel corpus is defined as
\begin{equation}
\mathcal{L}_{HsemaCL}^{\mathbb{Q}}
=
\mathcal{L}_{sentCL}^{\mathbb{Q}}
+
\lambda_t \mathcal{L}_{tokenCL}^{\mathbb{Q}}
\end{equation}
where $\lambda_t$ is the weight hyperparameter for the token-level loss.

Based on the supervision signals provided by the above parallel and non-parallel corpora, we define the hierarchical semantic contrastive loss as 
\begin{equation}
\mathcal{L}_{HsemaCL}
=
\mathcal{L}_{HsemaCL}^{\mathbb{P}}
+
\lambda_q \mathcal{L}_{HsemaCL}^{\mathbb{Q}}
\end{equation}

\subsection{Language Debiasing Contrastive Learning}
Sentences in the same language may express different semantics, but they often share similar lexical and syntactic patterns. In contrast, sentences from different languages should be distinguishable in the language-specific space. Therefore, we introduce a language debiasing contrastive loss to encourage the language component $\mathbf{z}_l$ to capture language-related information by pulling same-language representations closer and pushing different-language representations apart. Specifically, for each sample $x_k\in\mathcal{B}$, where $\mathcal{B}$ consists of samples from $\mathbb{P}$ and $\mathbb{Q}$, we denote its language as $\operatorname{lang}(x_k)$ and obtain its language-specific representation as
\begin{equation}
\mathbf{l}_k = W_l f(x_k).
\end{equation}
For the anchor sample $x_k$, samples satisfying $\operatorname{lang}(x_p)=\operatorname{lang}(x_k)$ are treated as positives, while the negative representation set is defined as 
$\mathcal{N}^{l}(x_k)=\{\mathbf{l}_r \mid x_r\in\mathcal{B}, 
\operatorname{lang}(x_r)\neq\operatorname{lang}(x_k)\}$. Based on these positive and negative samples, the language debiasing contrastive loss is formulated as
\begin{equation}
\mathcal{L}_{\mathrm{langCL}}
=
\frac{1}{|\mathcal{B}|}
\sum_{x_k\in\mathcal{B}}
\ell(\mathbf{l}_k,\mathbf{l}_p,\mathcal{N}^{l}(x_k))
\end{equation}

\begin{table*}[t]
\small
\renewcommand{\arraystretch}{1.23}
\centering
\resizebox{\textwidth}{!}{
\begin{tabular}{lccccccccccccccc}
\toprule
\textbf{Model} 
& \textbf{Avg} 
& \textbf{ar} 
& \textbf{de} 
& \textbf{en} 
& \textbf{es} 
& \textbf{fr} 
& \textbf{hi} 
& \textbf{id} 
& \textbf{it} 
& \textbf{ja} 
& \textbf{nl} 
& \textbf{pt} 
& \textbf{ru} 
& \textbf{vi} 
& \textbf{zh} \\
\midrule

\multicolumn{16}{l}{\textbf{Lexical Retrieval}} \\
\hdashline

BM25 
& 14.2 & 11.1 & 13.6 & 18.4 & 15.8 & 15.5 & 13.4 & 14.9 & 15.3 & 14.1 & 14.0 & 15.2 & 12.4 & 13.6 & 11.6 \\

\midrule

\multicolumn{16}{l}{\textbf{English Supervision}} \\
\hdashline

mDPR 
& 15.8 & 12.4 & 15.7 & 31.4 & 16.7 & 16.0 & 13.2 & 12.3 & 15.6 & 15.2 & 15.1 & 15.8 & 13.7 & 12.3 & 15.4 \\

mContriever 
& 17.7 & 16.1 & 23.2 & 29.5 & 15.9 & 15.8 & 14.6 & 17.7 & 16.2 & 14.7 & 19.7 & 18.4 & 18.1 & 13.7 & 14.4 \\

mColBERT 
& 18.3 & 16.8 & 24.9 & 30.4 & 16.1 & 14.6 & 17.2 & 14.9 & 18.7 & 17.1 & 20.2 & 19.2 & 17.7 & 13.9 & 14.2 \\

ColBERT-XM 
& 26.2 & 19.5 & 27.0 & \underline{37.2} & 28.5 & 26.9 & 23.8 & 26.3 & 26.5 & 24.1 & 27.5 & 27.6 & 25.1 & 22.6 & 24.6 \\

$\mathrm{mBERT}_{\mathrm{AGG}}$ 
& 24.9 & 18.2 & 25.3 & 33.6 & 27.9 & 26.5 & 22.6 & 25.7 & 26.9 & 22.2 & 25.8 & 26.7 & 24.3 & 19.4 & 23.1 \\

$\mathrm{XLM\mbox{-}R}_{\mathrm{AGG}}$ 
& 26.3 & 19.1 & 25.9 & 35.2 & 29.1 & 28.3 & 24.4 & 27.3 & 28.8 & 23.8 & 27.1 & \underline{28.3} & 25.4 & 21.2 & 24.8 \\

\midrule

\multicolumn{16}{l}{\textbf{Multilingual Supervision}} \\
\hdashline

$\mathrm{mE5}_{\mathrm{BASE}}$
& 26.5 & 20.5 & 27.1 & 35.0 & 28.9 & \underline{30.0} & 24.2 & 26.1 & 28.0 & 25.0 & 27.3 & 27.5 & 24.5 & 23.9 & 22.9 \\

BGE
& 27.1 & 20.8 & 27.7 & 36.5 & 29.2 & \textbf{30.1} & 25.1 & 27.4 & 28.8 & 25.3 & \underline{27.9} & 27.8 & 24.4 & \textbf{24.5} & 23.4 \\

$\mathrm{BGE}_{\mathrm{HN}}$
& \underline{27.4} & \underline{20.9} & \underline{27.9} & \underline{37.2} & \underline{29.3} & 29.9 & \underline{25.2} & \textbf{28.5} & \underline{29.1} & \textbf{25.8} & 27.4 & 27.7 & \underline{25.7} & \textbf{24.5} & \underline{25.1} \\

\midrule

\multicolumn{16}{l}{\textbf{Ours}} \\
\hdashline

$\mathrm{Ours}_{\mathrm{mBERT}}$ 
& 26.8$^\dagger$ 
& 19.3$^\dagger$ 
& 27.3$^\dagger$ 
& 36.7$^\dagger$ 
& 29.2$^\dagger$ 
& 28.3$^\dagger$ 
& 24.6$^\dagger$ 
& 27.8$^\dagger$ 
& 28.4$^\dagger$ 
& 24.5$^\dagger$ 
& 27.8$^\dagger$ 
& 28.2$^\dagger$ 
& 24.9$^\dagger$ 
& 22.9$^\dagger$ 
& \underline{25.1}$^\dagger$ \\

$\mathrm{Ours}_{\mathrm{XLM\mbox{-}R}}$ 
& \textbf{27.8}$^{\dagger\ddagger}$ 
& \textbf{21.1}$^{\dagger\ddagger}$ 
& \textbf{28.2}$^{\dagger\ddagger}$ 
& \textbf{37.4}$^{\dagger\ddagger}$ 
& \textbf{29.8}$^{\dagger\ddagger}$ 
& 29.7$^\dagger$ 
& \textbf{25.3}$^{\dagger\ddagger}$ 
& \underline{28.3}$^\dagger$ 
& \textbf{29.6}$^{\dagger\ddagger}$ 
& \underline{25.5}$^\dagger$ 
& \textbf{28.7}$^{\dagger\ddagger}$ 
& \textbf{28.6}$^{\dagger\ddagger}$ 
& \textbf{26.2}$^{\dagger\ddagger}$ 
& \underline{24.1}$^\dagger$ 
& \textbf{26.3}$^{\dagger\ddagger}$ \\

\bottomrule
\end{tabular}
}
\caption{In-domain zero-shot results on mMARCO using MRR@10.
$\dagger$ denotes significant improvements with paired t-test at $p<0.05$ over the baseline method using the same backbone model, and $\ddagger$ denotes significant improvements over the strongest baseline in the same column.
\textbf{Bold} and \underline{underline} indicate the best and second-best results, respectively.}
\label{tab:example-tabl1}
\end{table*}

\begin{table*}[t]
\centering
\renewcommand{\arraystretch}{1.45}
\setlength{\tabcolsep}{4.2pt}

\resizebox{\textwidth}{!}{
\begin{tabular}{l c cccccccccc c cccccccc c}
\toprule
\multirow{2}{*}{\textbf{Model}} 
& \multirow{2}{*}{\textbf{Avg}}
& \multicolumn{11}{c}{\textbf{With Parallel Data}}
& \multicolumn{9}{c}{\textbf{Without Parallel Data}} \\
\cmidrule(lr){3-13} \cmidrule(lr){14-22}
& 
& \textbf{ar} & \textbf{de} & \textbf{en} & \textbf{es} & \textbf{fr} 
& \textbf{hi} & \textbf{id} & \textbf{ja} & \textbf{ru} & \textbf{zh} 
& \textbf{Avg}$^{p}$
& \textbf{bn} & \textbf{fa} & \textbf{fi} & \textbf{ko} & \textbf{sw} 
& \textbf{te} & \textbf{th} & \textbf{yo} 
& \textbf{Avg}$^{np}$ \\
\midrule

\multicolumn{22}{l}{\textbf{Lexical Retrieval}} \\
\hdashline

BM25
& 38.5
& 48.1 & 22.6 & 35.1 & 31.9 & 18.3
& 45.8 & 44.9 & 36.9 & 33.4 & 18.0
& 33.5
& 50.8 & 33.3 & 55.1 & 41.9 & 38.3
& 49.4 & 48.4 & 40.6
& 44.7 \\

\midrule

\multicolumn{22}{l}{\textbf{English Supervision}} \\
\hdashline

mDPR
& 41.8
& 49.9 & 49.0 & \underline{39.4} & \textbf{47.8} & 43.5
& 38.3 & 27.2 & 43.9 & 40.7 & \textbf{51.2}
& 43.1
& 44.3 & \underline{48.0} & 47.2 & 41.9 & 29.9
& 35.6 & 35.8 & 39.6
& 40.3 \\

mContriever
& 43.1
& 52.5 & 40.8 & 36.4 & 41.8 & 31.4
& 28.6 & 39.2 & 42.4 & 39.1 & 41.0
& 39.3
& 50.1 & 21.5 & 60.2 & 48.3 & \underline{56.0}
& 52.8 & 51.7 & 41.5
& 47.8 \\

mColBERT
& 44.1
& 57.1 & 33.4 & 38.8 & 42.6 & 26.7
& 47.0 & 29.8 & 49.6 & \underline{47.7} & 39.8
& 41.3
& 54.6 & 46.0 & 46.5 & 48.7 & 35.8
& 46.2 & 48.1 & \underline{56.1}
& 47.8 \\

ColBERT-XM
& 52.0
& 60.4 & \underline{53.1} & 37.9 & 44.1 & 43.3
& 49.1 & 45.5 & \underline{54.6} & \underline{47.7} & 45.6
& 48.1
& 59.7 & 46.9 & 63.8 & 58.8 & 50.6
& 58.4 & 65.5 & 51.2
& 56.9 \\

$\mathrm{mBERT}_{\mathrm{AGG}}$
& 47.0
& 59.4 & 50.4 & 38.6 & 44.6 & 43.6
& 37.4 & 42.1 & 50.2 & 46.3 & 44.3
& 45.7
& 51.0 & 44.5 & 65.3 & 47.8 & 48.5
& 48.0 & 31.1 & 52.6
& 48.6 \\

$\mathrm{XLM\mbox{-}R}_{\mathrm{AGG}}$
& 52.3
& \underline{61.4} & 52.6 & 38.4 & 42.9 & 41.3
& 46.2 & \textbf{48.4} & 53.9 & 46.5 & 41.0
& 47.3
& 61.4 & 46.5 & 66.2 & 57.9 & 47.5
& \underline{71.2} & 66.8 & 51.2
& 58.6 \\

\midrule

\multicolumn{22}{l}{\textbf{Ours}} \\
\hdashline

$\mathrm{Ours}_{\mathrm{mBERT}}$
& \underline{53.8}$^\dagger$
& 61.3 & \underline{53.1} & 38.8 & 45.4 & \underline{44.3}$^\dagger$
& \underline{49.4}$^\dagger$ & 46.9 & 54.2 & 47.4 & 46.3
& \underline{48.7}$^\dagger$
& \underline{61.6}$^\dagger$ & 47.5 & \underline{66.6}$^\dagger$ & \underline{59.8}$^\dagger$ & 53.9
& 68.9 & \underline{67.6}$^\dagger$ & 54.7
& \underline{60.1}$^\dagger$ \\

$\mathrm{Ours}_{\mathrm{XLM\mbox{-}R}}$
& \textbf{55.0}$^\dagger$
& \textbf{62.1}$^\dagger$ & \textbf{53.6}$^\dagger$ & \textbf{39.5}$^\dagger$ & \underline{46.7} & \textbf{45.1}$^\dagger$
& \textbf{50.2}$^\dagger$ & \underline{47.8} & \textbf{55.4}$^\dagger$ & \textbf{48.6}$^\dagger$ & \underline{48.8}
& \textbf{49.8}$^\dagger$
& \textbf{62.5}$^\dagger$ & \textbf{48.3}$^\dagger$ & \textbf{67.7}$^\dagger$ & \textbf{60.6}$^\dagger$ & \textbf{56.2}$^\dagger$
& \textbf{71.4}$^\dagger$ & \textbf{68.4}$^\dagger$ & \textbf{57.5}$^\dagger$
& \textbf{61.6}$^\dagger$ \\

\bottomrule
\end{tabular}
}

\caption{Out-of-domain zero-shot results on MIRACL using nDCG@10. Languages are grouped according to whether their corresponding English–X parallel data are used during training.
$\dagger$ denotes significant improvements with paired t-test at $p<0.05$ over the strongest baseline in the same column. \textbf{Bold} and \underline{underline} indicate the best and second-best results.}
\label{tab:example-tabl2}
\end{table*}

\subsection{Joint Training}
To optimize the disentangler, we combine hierarchical semantic contrastive
learning, language debiasing contrastive learning, and the orthogonal constraint. The disentangled objective is defined as
\begin{equation}
\mathcal{L}_{dec}
=
\mathcal{L}_{HsemaCL}
+
\mathcal{L}_{langCL}
+
\lambda_o \mathcal{L}_{orth}
\end{equation}
where $\lambda_o$ controls the contribution of the orthogonal constraint.
Overall, we jointly optimize the retrieval objective and the disentanglement objective. Given a query $q^l$, its positive document $d_q^+$, and the negative document set $\mathcal{D}_q^-$, we define the retrieval representation as
$r(x)=W_s f(x)$. The overall training objective is formulated as
\begin{equation}
\mathcal{L}
=
\ell\left(
r\left(q^{(\ell)}\right),
r\left(d_q^{+}\right),
\left\{r(d): d \in D_q^{-}\right\}
\right)
+
\lambda_d \mathcal{L}_{\mathrm{dec}}
\end{equation}
where $\lambda_d$ controls the contribution of $\mathcal{L}_{\mathrm{dec}}$.
The whole procedure is described in the supplementary materials E.

\begin{table}[!t]
\centering
\small
\setlength{\tabcolsep}{4.2pt}
\renewcommand{\arraystretch}{0.8}
\resizebox{0.95\columnwidth}{!}{%
\begin{tabular}{lcccc}
\toprule
\textbf{Model} 
& \textbf{MRR@10} 
& \textbf{\boldmath$\Delta$} 
& \textbf{Recall@1k} 
& \textbf{\boldmath$\Delta$} \\
\midrule
XLM-R 
& 22.6 & -- 
& 80.9 & -- \\
\midrule
\quad w/ Dis. 
& 24.9 & $+2.3$ 
& 84.1 & $+3.2$ \\
\quad w/ HSA 
& 23.1 & $+0.5$ 
& 81.3 & $+0.4$ \\
\quad w/ LDCL 
& 22.9 & $+0.3$ 
& 81.2 & $+0.3$ \\
\midrule
\quad w/ Dis. + HSA 
& 26.7 & $+4.1$ 
& 86.2 & $+5.3$ \\
\quad w/ Dis. + LDCL 
& 25.4 & $+2.8$ 
& 85.5 & $+4.6$ \\
\quad w/ HSA + LDCL 
& 23.5 & $+0.9$ 
& 81.6 & $+0.7$ \\
\midrule
Full Model
& 27.8 & \textbf{$+5.2$} 
& 87.1 & \textbf{$+6.2$} \\
\bottomrule
\end{tabular}%
}
\caption{
Ablation on three components of the disentanglement method on mMARCO. i) Dis.: Language-semantic Disentangler, ii) HSA: Hierarchical Semantic Alignment, and iii) LDCL: Language Debiasing Contrastive Learning.}
\label{tab:ablation_disentanglement}
\end{table}

\section{Experiments}
\subsection{Experimental Setup}
\noindent\textbf{Datasets and Metrics.}
For multilingual dense retrieval, we use only the English~(MS MARCO~\cite{bajaj2016ms}) passage ranking dataset as retrieval supervision for model training. For cross-lingual semantic alignment, we use the WikiMatrix~\cite{schwenk2021wikimatrix} dataset. Specifically, we select parallel sentences between English and the 13 non-English languages covered by mMARCO~\cite{bonifacio2021mmarco}. To evaluate zero-shot multilingual retrieval performance, we use the mMARCO development set for in-domain evaluation and the MIRACL development set for out-of-domain evaluation. Following previous studies~\cite{bonifacio2021mmarco,zhang2023miracl}, we report MRR@10 and Recall@1k on mMARCO, and nDCG@10 and Recall@100 on MIRACL. 

\noindent\textbf{Implementation Details.}
We implement the multilingual dense retriever based on Tevatron~\cite{gao2022tevatron}.
For multilingual dense retriever training, we set the learning rate to 3e-6 and use the Adam optimizer for 4 epochs, corresponding to 60K steps. The batch size is set to 24. For each query, we sample one positive sample and seven negative samples, while other passages in the same batch are used as in-batch negatives. The maximum input lengths for queries and passages are set to 64 and 256, respectively. During inference, the maximum input lengths for queries and passages are set to 64 and 256 on mMARCO, and 128 and 256 on MIRACL. The loss weights $\lambda_d$, $\lambda_o$, $\lambda_q$, and $\lambda_t$ are set to 0.05, 0.2, 0.1, and 0.2, respectively, based on a simple grid search.

\noindent\textbf{Baseline Methods.}
To provide a comprehensive comparison, we consider three types of baselines: lexical retrieval, English-supervised retrieval, and multilingual-supervised retrieval. The lexical baseline is BM25~\cite{robertson2009probabilistic}. 
The English-supervised baselines include mDPR~\cite{zhang2023miracl}, mContriever~\cite{izacard2021unsupervised}, mColBERT~\cite{khattab2020colbert}, ColBERT-XM~\cite{louis2025colbert}, mBERT$_{\mathrm{AGG}}$~\cite{lin2023maggretriever}, and XLM-R$_{\mathrm{AGG}}$~\cite{lin2023maggretriever}. 
The multilingual-supervised baselines include $\mathrm{mE5}_{\mathrm{BASE}}$~\cite{wang2022text}, BGE~\cite{chen2024bge}, and $\mathrm{BGE}_{\mathrm{HN}}$~\cite{huang2025boosting}. 
Notably, the multilingual-supervised baselines are trained with much larger multilingual retrieval data and are included as reference systems. For MIRACL, multilingual-supervised models using MIRACL training data are excluded to maintain the out-of-domain zero-shot evaluation setting. 

\subsection{Overall Performance}
\noindent \textbf{In-domain Performance.} As shown in Table~\ref{tab:example-tabl1}, our method outperforms the baselines on the mMARCO dataset with 14 languages. Specifically, our method achieves absolute gains of 1.5 and 0.4 MRR@10 points over the strongest English-supervised and multilingual-supervised baselines in terms of average scores, respectively. The superior performance can be attributed to two aspects: (1) the disentangler separates retrieval-relevant semantic features from linguistic information, enabling the model to perform semantic matching in a cleaner semantic space. (2) hierarchical semantic alignment improves fine-grained cross-lingual semantic matching. Furthermore, the consistent gains across different backbone models demonstrate the effectiveness and generalization ability of our method for zero-shot multilingual retrieval.

\noindent \textbf{Out-of-domain Performance.}
As shown in Table~\ref{tab:example-tabl2}, we evaluate the out-of-domain zero-shot cross-lingual retrieval capability of the model on the MIRACL dataset. Compared to baseline methods, our method achieves optimal performance on most languages.
Specifically, our method surpasses the strongest baseline XLM-R$_{\mathrm{AGG}}$ by 2.7 nDCG@10 points on average, demonstrating its robustness under out-of-domain retrieval settings. It also achieves the best average performance on both languages with and without parallel corpora. The larger improvement on languages without parallel corpora suggests that the proposed disentanglement and semantic consistency objectives facilitate indirect cross-lingual transfer, enabling the model to learn more stable language-agnostic retrieval representations.

\begin{table}[!t]
\centering
\renewcommand{\arraystretch}{0.8}
\setlength{\tabcolsep}{4.5pt}
\resizebox{0.95\columnwidth}{!}{%
\begin{tabular}{l l c c}
\toprule
\textbf{Setting} & \textbf{Configuration} & \textbf{MRR@10} & \textbf{Recall@1k} \\
\midrule
\textbf{\hspace{0.4em}S1} & Random language pairs & 27.4 & 86.7 \\
\textbf{\hspace{0.4em}S2} & English-centric pairs & 27.8 & 87.1 \\
\textbf{\hspace{0.4em}S3} & Hindi-centric pairs & 27.1 & 86.3 \\
\textbf{\hspace{0.4em}S4} & Intra-family pairs & 27.6 & 86.9 \\
\textbf{\hspace{0.4em}S5} & High--low-resource pairs & 27.4 & 86.6 \\
\bottomrule
\end{tabular}
}
\caption{Experimental results of different parallel sentence pair construction strategies with average scores. }
\label{tab:example-tabl5}
\end{table}

\begin{figure}[!t]
    \centering
    \includegraphics[width=0.9\linewidth]{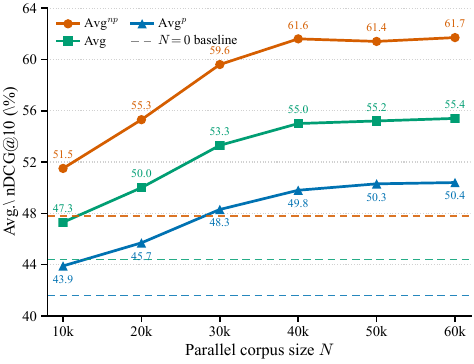}
    \caption{Performance on MIRACL with average nDCG@10 under different parallel-corpus sizes $N$. Solid lines indicate the averages over all languages (Avg), languages with parallel data ($\mathrm{Avg}^{\mathrm{p}}$), and without ($\mathrm{Avg}^{\mathrm{np}}$). Dashed lines indicate the corresponding $N{=}0$ baselines.} 
    \label{fig4}
\end{figure}

\begin{figure*}[!t]
    \centering
    \includegraphics[width=0.98\textwidth]{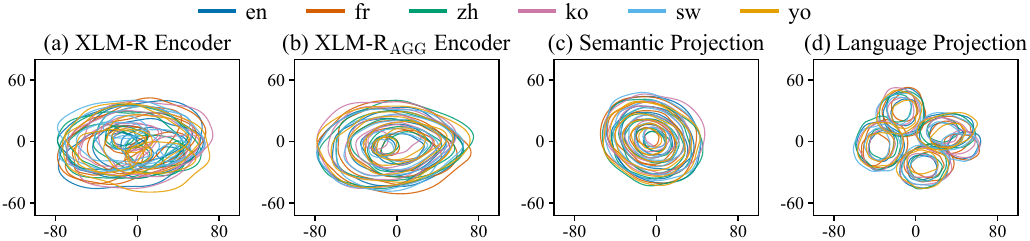}
    \caption{t-SNE visualization of different representation spaces. The subfigure titles identify the model modules, whose outputs are visualized after dimensionality reduction. Languages mix in (c) but separate in (d), indicating successful disentanglement.}
    \label{fig:fig2}
\end{figure*}

\begin{table}[!t]
\centering
\small
\setlength{\tabcolsep}{4.2pt}
\renewcommand{\arraystretch}{0.8}
\begin{tabular}{llcc}
\toprule
\textbf{Resource Group} & \textbf{Language} & \textbf{XLM-R$_{\mathrm{AGG}}$} & \textbf{Ours} \\
\midrule
High-resource & es, fr, pt, ru, de & 27.4 & 28.5 \\
Medium-resource & vi, id, ar, ja & 22.9 & 24.8 \\
Low-resource & nl, zh, hi, it & 26.3 & 27.5 \\
\bottomrule
\end{tabular}
\caption{
Resource-level performance comparison on mMARCO in terms of average MRR@10.
Resource here refers to the number of EN–X parallel pairs in WikiMatrix, not the general resource level of the language.}
\label{tab:resource_breakdown_mmarco}
\end{table}

\begin{table}[!t]
\centering
\small
\setlength{\tabcolsep}{4.2pt}
\renewcommand{\arraystretch}{0.8}
\begin{tabular}{llcc}
\toprule
\textbf{Language Family} & \textbf{Language} & \textbf{XLM-R$_{\mathrm{AGG}}$} & \textbf{Ours} \\
\midrule
Germanic & en, de, nl & 29.4 & 31.4 \\
Romance & es, fr, it, pt & 28.6 & 29.4 \\
Indo-Iranian & hi & 24.4 & 25.3 \\
\bottomrule
\end{tabular}
\caption{
Model performance comparison with average MRR@10 across different language families on mMARCO.
}
\label{tab:family_breakdown_mmarco}
\end{table}

\subsection{Ablation Studies}
As shown in Table~\ref{tab:ablation_disentanglement}, we investigate the effectiveness of each component within our method on mMARCO. Performance gains are observed when each component is used independently or in combination, indicating that each component makes a positive contribution to the multilingual dense retriever.
In particular, the language-semantic disentangler improves MRR@10 by 2.3 points over the base model, and further combination with other components achieves consistent performance gains, showing that separating semantic features from linguistic information effectively alleviates language bias. On top of the disentangler, hierarchical semantic alignment (HSA) and language debiasing contrastive learning (LDCL) can improve MRR@10 by 1.8 and 0.5 points, respectively. The larger gain from HSA confirms that explicit cross-lingual semantic alignment contributes most directly to retrieval, while LDCL plays a complementary role by constraining the language component to absorb linguistic features and indirectly purifying the semantic space.

\section{Analysis and Discussion}
\subsection{Impact of Parallel Supervision}
\noindent \textbf{Connection Patterns.}
 As shown in Table~\ref{tab:example-tabl5}, we first examine the cross-lingual alignment with different parallel-pair construction strategies. The results show that the English-centric pairs (\textbf{S2}) perform best, followed by intra-family pairs (\textbf{S4}), while random, Hindi-centric, and high--low-resource pairs lead to slightly lower results. The overall gaps are minor, suggesting that our framework is relatively robust to the connection pattern of parallel pairs. The advantage of English-centric pairs is expected, since retrieval supervision is learned from English and can be transferred more directly through English-centered alignment.

\noindent \textbf{Supervision Scale.}
 As shown in Figure~\ref{fig4}, we further vary the number of sampled parallel sentence pairs to study the effect of supervision scale. The results show that the performance improves steadily from 10K to 60K parallel pairs, with larger gains before 40K and only marginal improvements afterward. This indicates that parallel supervision helps build a shared semantic space, but its benefit saturates after a moderate data scale. Similar trends across languages with and without direct parallel data suggest that the learned alignment signals also benefit indirectly connected languages.

\subsection{Impact of Language Families and Resource Levels}
To examine whether our method generalizes across diverse languages, we further analyze its performance from two perspectives: resource level and language family. Table~\ref{tab:resource_breakdown_mmarco} shows consistent improvements over XLM-R$_{\mathrm{AGG}}$ across all language groups. This indicates that our method remains effective under varying amounts of parallel alignment resources, rather than relying only on resource-rich languages. Moreover, Table~\ref{tab:family_breakdown_mmarco} shows that our method consistently outperforms XLM-R$_{\mathrm{AGG}}$ across selected language families, indicating that our method can generalize well across different language families. Further details are provided in the supplementary materials G.

\subsection{Visualization}
To examine the encoder representation space, we visualize sentence embeddings via t-SNE in Figure~\ref{fig:fig2}(a)-(d). The XLM-R encoder still exhibits visible language-wise separation, indicating that semantic and linguistic information remain entangled in the original multilingual representations. XLM-R$_{\mathrm{AGG}}$ partially mitigates this, increasing cross-lingual overlap but failing to fully eliminate the language-induced distribution shift. In contrast, our semantic projections collapse cross-lingual texts into a unified, highly overlapping cluster, while the language projections retain distinct language-specific patterns, confirming the effectiveness of our disentanglement.

\section{Conclusion}
In this work, we propose a method to boost cross-lingual transfer for multilingual dense retrieval. By separating semantic and linguistic features and jointly optimizing disentangled objectives with the retrieval objective, our method learns more transferable semantic representations. Extensive results on multilingual retrieval benchmarks show that our method consistently outperforms strong baselines, demonstrating its effectiveness in cross-lingual transfer.




\bibliography{aaai2027}


\clearpage
\appendix

\section*{Appendix}
\addcontentsline{toc}{section}{Appendix A}  

\setcounter{section}{0}  

\section{Dataset Statistics}
\label{A}
\begin{figure*}
    \centering
     \includegraphics[width=1\linewidth]{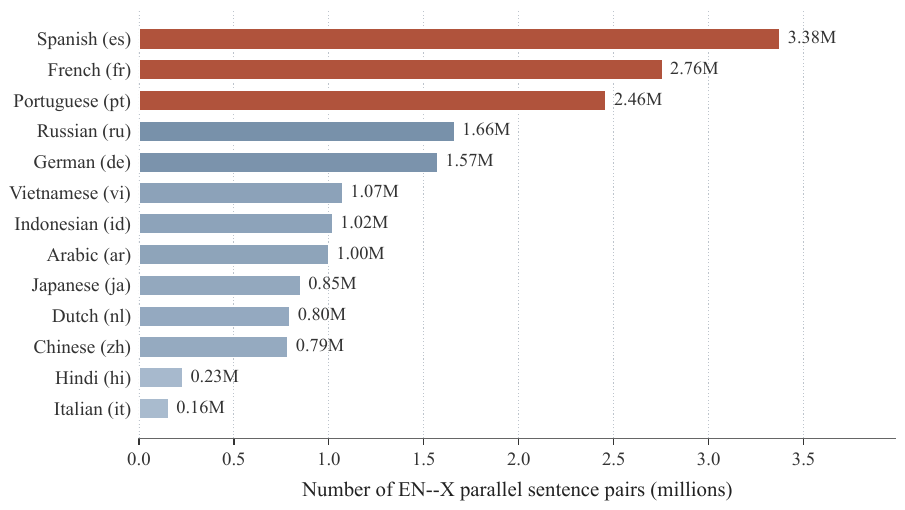}
    \caption{Number of EN–X parallel sentence pairs in WikiMatrix across the 13 non-English languages used for cross-lingual alignment training.}
    \label{fig:f4}
\end{figure*}

We use the mMARCO, WikiMatrix, and MIRACL datasets in our experiments. Their detailed information is provided as follows:
\begin{itemize}
    \item mMARCO~\cite{bonifacio2021mmarco}: a translated version of the MS MARCO dataset covering 14 languages, with each language containing approximately 8.8M passages and 500K training queries. 

    \item WikiMatrix~\cite{schwenk2021wikimatrix}: a large-scale multilingual parallel sentence dataset mined from Wikipedia across many language pairs, providing parallel sentence pairs for 85 languages. We sample parallel sentences between English and the other 13 languages in mMARCO, and the statistics of the parallel sentences are shown in Figure~\ref{fig:f4}.

    \item MIRACL~\cite{zhang2023miracl}: a multilingual retrieval benchmark dataset covering 18 languages, with 726k manual relevance judgments and a collection size of over 100 million documents. Each query is provided with an average of 10 manually verified relevance labels. The statistics of the MIRACL dataset are shown in Table~\ref{tab:fulu1}. 
\end{itemize}

\begin{table*}[!t]
\centering
\scriptsize
\renewcommand{\arraystretch}{1.25}
\setlength{\tabcolsep}{3.8pt}

\resizebox{\textwidth}{!}{
\begin{tabular}{llrrrrrrrrrr}
\toprule
\textbf{Lang} &  & \textbf{All} 
& \multicolumn{1}{c}{\textbf{Arabic}} 
& \multicolumn{1}{c}{\textbf{Bengali}} 
& \multicolumn{1}{c}{\textbf{English}} 
& \multicolumn{1}{c}{\textbf{Finnish}} 
& \multicolumn{1}{c}{\textbf{Indonesian}} 
& \multicolumn{1}{c}{\textbf{Japanese}} 
& \multicolumn{1}{c}{\textbf{Korean}} 
& \multicolumn{1}{c}{\textbf{Russian}} 
& \multicolumn{1}{c}{\textbf{Swahili}} \\
\midrule
\textbf{ISO} &  &  
& \multicolumn{1}{c}{\textbf{ar}} 
& \multicolumn{1}{c}{\textbf{bn}} 
& \multicolumn{1}{c}{\textbf{en}} 
& \multicolumn{1}{c}{\textbf{fi}} 
& \multicolumn{1}{c}{\textbf{id}} 
& \multicolumn{1}{c}{\textbf{ja}} 
& \multicolumn{1}{c}{\textbf{ko}} 
& \multicolumn{1}{c}{\textbf{ru}} 
& \multicolumn{1}{c}{\textbf{sw}} \\
\midrule
\multirow{2}{*}{\textbf{Train}}
& \#Q & 40{,}203 & 3{,}495 & 1{,}631 & 2{,}863 & 2{,}897 & 4{,}071 & 3{,}477 & 868 & 4{,}683 & 1{,}901 \\
& \#J & 343{,}177 & 25{,}382 & 16{,}754 & 29{,}416 & 20{,}350 & 41{,}358 & 34{,}387 & 12{,}767 & 33{,}921 & 9{,}359 \\
\midrule
\multirow{2}{*}{\textbf{Test}}
& \#Q & 13{,}495 & 2{,}896 & 411 & 799 & 1{,}271 & 960 & 860 & 213 & 1{,}252 & 482 \\
& \#J & 130{,}408 & 29{,}197 & 4{,}206 & 8{,}350 & 12{,}008 & 9{,}668 & 8{,}354 & 3{,}057 & 13{,}100 & 5{,}092 \\
\midrule
\textbf{Passages}
&  & 106{,}332{,}152 & 2{,}061{,}414 & 297{,}265 & 32{,}893{,}221 & 1{,}883{,}509 & 1{,}446{,}315 & 6{,}953{,}614 & 1{,}486{,}752 & 9{,}543{,}918 & 131{,}924 \\
\bottomrule
\end{tabular}
}

\vspace{0.8em}

\resizebox{\textwidth}{!}{
\begin{tabular}{llrrrrrrrrrr}
\toprule
\textbf{Lang} &  & \textbf{All} 
& \multicolumn{1}{c}{\textbf{Telugu}} 
& \multicolumn{1}{c}{\textbf{Thai}} 
& \multicolumn{1}{c}{\textbf{Spanish}} 
& \multicolumn{1}{c}{\textbf{Persian}} 
& \multicolumn{1}{c}{\textbf{French}} 
& \multicolumn{1}{c}{\textbf{Hindi}} 
& \multicolumn{1}{c}{\textbf{Chinese}} 
& \multicolumn{1}{c}{\textbf{German}} 
& \multicolumn{1}{c}{\textbf{Yoruba}} \\
\midrule
\textbf{ISO} &  &  
& \multicolumn{1}{c}{\textbf{te}} 
& \multicolumn{1}{c}{\textbf{th}} 
& \multicolumn{1}{c}{\textbf{es}} 
& \multicolumn{1}{c}{\textbf{fa}} 
& \multicolumn{1}{c}{\textbf{fr}} 
& \multicolumn{1}{c}{\textbf{hi}} 
& \multicolumn{1}{c}{\textbf{zh}} 
& \multicolumn{1}{c}{\textbf{de}} 
& \multicolumn{1}{c}{\textbf{yo}} \\
\midrule
\multirow{2}{*}{\textbf{Train}}
& \#Q & 40{,}203 & 3{,}452 & 2{,}972 & 2{,}162 & 2{,}107 & 1{,}143 & 1{,}169 & 1{,}312 
& \multicolumn{1}{c}{--} & \multicolumn{1}{c}{--} \\
& \#J & 343{,}177 & 18{,}608 & 21{,}293 & 21{,}531 & 21{,}844 & 11{,}426 & 11{,}668 & 13{,}113 
& \multicolumn{1}{c}{--} & \multicolumn{1}{c}{--} \\
\midrule
\multirow{2}{*}{\textbf{Test}}
& \#Q & 13{,}495 & 828 & 733 & 648 & 632 & 343 & 350 & 393 & 305 & 119 \\
& \#J & 130{,}408 & 1{,}606 & 7{,}573 & 6{,}443 & 6{,}571 & 3{,}429 & 3{,}494 & 3{,}928 & 3{,}144 & 1{,}188 \\
\midrule
\textbf{Passages}
&  & 106{,}332{,}152 & 518{,}079 & 542{,}166 & 10{,}373{,}953 & 2{,}207{,}172 & 14{,}636{,}953 & 506{,}264 & 4{,}934{,}368 & 15{,}866{,}222 & 49{,}043 \\
\bottomrule
\end{tabular}
}

\caption{Statistics of MIRACL. The \#Q and \#J denote the numbers of queries and relevance judgments, respectively.}
\label{tab:fulu1}
\end{table*}

\section{Implementation Details}
\label{B}
We implement all models by Pytorch~\cite{paszke2019pytorch} and Huggingface's Transformers library~\cite{wolf2019huggingface}. All experiments are conducted on an Nvidia A100 80G GPU. Note that for languages included in mMARCO, the number of parallel sentence pairs with English in WikiMatrix ranges from 157,000 to 3,377,000. During training, we sample 40,000 parallel sentence pairs for each language. Moreover, we construct $\mathbb{Q}$ from Wikipedia monolingual texts in the MIRACL languages that are not covered by the parallel alignment data, including Bengali, Persian, Finnish, Korean, Swahili, Telugu, Thai, and Yoruba. These sentences are used without parallel counterparts, retrieval labels, or relevance judgments. In each batch, we sample $N$ parallel sentence pairs from $\mathbb{P}$ and $N$ monolingual sentences from $\mathbb{Q}$. This design enables the model to learn direct cross-lingual alignment from $\mathbb{P}$ and indirect semantic consistency from $\mathbb{Q}$, without using any target-language retrieval supervision. Each reported result is averaged over three independent runs.


\section{Baseline Details}
\label{C}
We provide a more detailed introduction to the following baselines used for comparison:
\begin{itemize}
    \item \textbf{BM25}~\cite{robertson2009probabilistic}: a retrieval model based on lexical matching.

    \item \textbf{mDPR}~\cite{zhang2023miracl}: a single-vector dense retrieval model initialized by mBERT and fine-tuned on MS MARCO using contrastive learning.

    \item \textbf{mColBERT}~\cite{khattab2020colbert}: a multi-vector dense retrieval model implemented similarly to mDPR, but using multiple word-level vectors for fine-grained similarity matching.

    \item \textbf{mContriever}~\cite{izacard2021unsupervised}: a single-vector dense retrieval model initialized from mBERT and further fine-tuned on MS MARCO.

    \item \textbf{ColBERT-XM}~\cite{louis2025colbert}: a modular multi-vector dense retrieval model fine-tuned on MS MARCO based on the XMOD framework.

    \item \textbf{mBERT$_{\mathbf{AGG}}$}~\cite{lin2023maggretriever}: a dense retrieval model using semantic and lexical features in mBERT to achieve zero-shot transfer. 

    \item \textbf{XLM-R$_{\mathbf{AGG}}$}~\cite{lin2023maggretriever}: Unlike mBERT$_{\mathrm{AGG}}$, this baseline uses XLM-R as the backbone model for fine-tuning on MS MARCO. 

    \item \textbf{mE5$_{\mathbf{BASE}}$}~\cite{wang2022text}: a multilingual embedding model designed for retrieval and semantic matching, which leverages large-scale corpora for weakly supervised contrastive pre-training and large-scale retrieval-labeled data for supervised fine-tuning. 

    \item \textbf{BGE}~\cite{chen2024bge}: a multilingual embedding model for multilingual and cross-lingual retrieval, pre-trained on large-scale multilingual unsupervised data and fine-tuned on retrieval-labeled data. 

    \item \textbf{BGE$_{\mathbf{HN}}$}~\cite{huang2025boosting}: a multilingual dense retriever initialized with BGE and further optimized using the constructed hard negative set and an efficient mini-batch sampling strategy. 
\end{itemize}

\section{Sensitivity Analysis of Loss Weight Hyperparameters}
\label{D}
Table~\ref{tab:loss_weight_sensitivity} reports the retrieval performance under different loss weight settings. To analyze the sensitivity of our model to key loss weights, we vary $\lambda_d$, $\lambda_o$, $\lambda_q$, and $\lambda_t$ from their default values and compare the corresponding nDCG@10 scores. We observe that individually increasing $\lambda_d$ or $\lambda_o$, or decreasing $\lambda_q$ or $\lambda_t$, leads to performance drops of 0.5, 0.7, 0.4, and 0.5 percentage points, respectively. This suggests that overly strong disentanglement or orthogonal constraints may interfere with the optimization of the original retrieval objective. Meanwhile, weakening the non-parallel semantic consistency loss or the token-level semantic alignment loss reduces the model's ability to handle languages without parallel data and fine-grained semantic units.
The performance degradation becomes more pronounced when multiple hyperparameters deviate from their default values simultaneously, e.g., $\lambda_d=0.1$, $\lambda_o=0.1$, $\lambda_q=0.2$, and $\lambda_t=0.3$. This indicates that different loss terms impose mutual constraints on each other, and overemphasizing a certain type of training signal may disrupt the balance among retrieval optimization, semantic alignment, and disentanglement regularization.




\begin{table}[!t]
\centering
\small
\setlength{\tabcolsep}{4.2pt}
\renewcommand{\arraystretch}{1.12}
\begin{tabular*}{0.87\columnwidth}{@{\extracolsep{\fill}}cccccc@{}}
\toprule
\textbf{\boldmath$\lambda_d$} 
& \textbf{\boldmath$\lambda_o$}
& \textbf{\boldmath$\lambda_q$}
& \textbf{\boldmath$\lambda_t$}
& \textbf{nDCG@10} 
& \textbf{\boldmath$\Delta$} \\
\midrule
0.05 & 0.20 & 0.10 & 0.20 
& \textbf{55.0} & -- \\
\midrule
0.10 & 0.20 & 0.10 & 0.20 
& 54.5 & $-0.5$ \\
0.05 & 0.30 & 0.10 & 0.20 
& 54.3 & $-0.7$ \\
0.05 & 0.20 & 0.05 & 0.20 
& 54.6 & $-0.4$ \\
0.05 & 0.20 & 0.10 & 0.10
& 54.5 & $-0.5$ \\
0.03 & 0.10 & 0.10 & 0.20 
& 54.1 & $-0.9$ \\
0.05 & 0.20 & 0.20 & 0.10
& 54.6 & $-0.4$ \\
0.10 & 0.10 & 0.20 & 0.30
& 53.8 & $-1.2$ \\
\bottomrule
\end{tabular*}
\caption{
Sensitivity analysis of loss weights on MIRACL.
The first row denotes the default configuration.
$\Delta$ denotes the absolute change in nDCG@10 compared with the default setting.
}
\label{tab:loss_weight_sensitivity}
\end{table}

\section{Algorithm Details}
\label{E}
Algorithm~\ref{alg:hdcl} summarizes the training procedure of our method. 
At each step, the model jointly optimizes the retrieval loss and the disentanglement loss.
The latter consists of hierarchical semantic alignment over parallel and non-parallel data, language debiasing contrastive learning, and the orthogonal constraint between semantic and language projection heads.

\begin{algorithm}[t]
\footnotesize
\caption{Disentangled Contrastive Learning}
\label{alg:hdcl}

\noindent\textbf{Input:}
Retrieval data $\mathcal{D}_{IR}$, parallel corpus $\mathbb{P}$, non-parallel corpus $\mathbb{Q}$, encoder $f_{\theta}$, semantic projector $W_s$, and language projector $W_l$.\\
\noindent\textbf{Output:}
Optimized multilingual dense retriever $f_{\theta}$.

\begin{algorithmic}[1]
\While{not converged}
    \State Sample batches $\mathcal{B}_{IR}$, $\mathcal{B}_{\mathbb{P}}$, and $\mathcal{B}_{\mathbb{Q}}$

    \For{each $(q_i,d_i^+)\in\mathcal{B}_{IR}$}
        \State Encode $q_i$, $d_i^+$ and in-batch negatives
    \EndFor
    \State Compute retrieval loss $\mathcal{L}_{IR}$

    \State \hspace*{-\algorithmicindent}\textit{Perform hierarchical semantic alignment.}
    \For{each sampled sentence instance}
        \If{parallel pair $(x_i,x_j)$ exists}
            \State $\mathbf{s}_i = W_s f_{\theta}(x_i),\quad \mathbf{s}_j = W_s f_{\theta}(x_j)$
            \State Contrast sentence pair and aligned tokens
            \State Accumulate $\mathcal{L}^{\mathbb{P}}_{sentCL}$ and $\mathcal{L}^{\mathbb{P}}_{tokenCL}$
        \Else
            \State Apply sentence- and token-level consistency
            \State Accumulate $\mathcal{L}^{\mathbb{Q}}_{sentCL}$ and $\mathcal{L}^{\mathbb{Q}}_{tokenCL}$
        \EndIf
    \EndFor
    \State $\mathcal{L}^{\mathbb{P}}_{HsemaCL}\leftarrow\mathcal{L}^{\mathbb{P}}_{sentCL}+\lambda_t\mathcal{L}^{\mathbb{P}}_{tokenCL}$
    \State $\mathcal{L}^{\mathbb{Q}}_{HsemaCL}\leftarrow\mathcal{L}^{\mathbb{Q}}_{sentCL}+\lambda_t\mathcal{L}^{\mathbb{Q}}_{tokenCL}$
    
    \State \hspace*{-\algorithmicindent}\textit{Perform language debiasing contrastive learning}
    \For{each $x_k\in\mathcal{B}_{\mathbb{P}}\cup\mathcal{B}_{\mathbb{Q}}$}
        \State $\mathbf{l}_k = W_l f_{\theta}(x_k)$
        \State \parbox[t]{0.78\linewidth}{Contrast $\mathbf{l}_k$ with same-language positives and different-language negatives}
    \EndFor
    \State Compute $\mathcal{L}_{langCL}$

    \State \hspace*{-\algorithmicindent}\textit{Construct disentangled objective}
    \For{each sentence representation $\mathbf{z}_k$ in the batch $\mathcal{B}$}
    \State $\mathbf{z}_{s,k} \leftarrow W_s\mathbf{z}_k,\quad
    \mathbf{z}_{l,k} \leftarrow W_l\mathbf{z}_k$
    \EndFor
    \State Stack $\{\mathbf{z}_{s,k}\}_{k=1}^{|\mathcal{B}|}$ and
    $\{\mathbf{z}_{l,k}\}_{k=1}^{|\mathcal{B}|}$ into
    $\mathbf{Z}_s$ and $\mathbf{Z}_l$
    \State $\mathcal{L}_{orth}\leftarrow\|\frac{1}{|\mathcal{B}|} \mathbf{Z}_s^{\top}\mathbf{Z}_l\|_F^2$
    \State $\mathcal{L}_{dec}\leftarrow
    \mathcal{L}_{HsemaCL}+\mathcal{L}_{langCL}+\lambda_o\mathcal{L}_{orth}$

    \State \hspace*{-\algorithmicindent}\textit{Perform joint optimization.}
    \State $\mathcal{L}\leftarrow\mathcal{L}_{IR}+\lambda_d\mathcal{L}_{dec}$
    \State Update $\theta$, $W_s$, and $W_l$ by minimizing $\mathcal{L}$
\EndWhile

\end{algorithmic}
\end{algorithm}

\begin{table}[t]
\centering
\small
\setlength{\tabcolsep}{10pt}
\renewcommand{\arraystretch}{1.15}

\begin{tabular}{lcc}   
\toprule
\textbf{Model}
& \textbf{MRR@10}
& \textbf{Recall@1k} \\
\midrule
Full Model
& 27.8 
& 87.1  \\
\quad w/o sentCL
& 26.6 
& 85.9  \\
\quad w/o tokenCL
& 27.1 
& 86.4 \\
\bottomrule
\end{tabular}
\caption{
Ablation on hierarchical semantic alignment with average scores on mMARCO.
}
\label{tab:ablation_hsa}
\end{table}


\begin{table}[!t]
\centering
\small
\setlength{\tabcolsep}{4.2pt}
\renewcommand{\arraystretch}{1.12}
\begin{tabular}{llcc}
\toprule
\textbf{Language Family} & \textbf{Language} & \textbf{XLM-R$_{\mathrm{AGG}}$} & \textbf{Ours} \\
\midrule
Germanic & en, de, nl & 29.4 & 31.4 \\
Romance & es, fr, it, pt & 28.6 & 29.4 \\
Indo-Iranian & hi & 24.4 & 25.3 \\
Slavic & ru & 25.4 & 26.2 \\ 
Semitic & ar & 19.1 & 21.1 \\
Sino-Tibetan & zh & 24.8 & 26.3 \\
Japonic & ja & 23.8 & 25.5 \\
Austronesian & id & 27.3 & 28.3 \\
Austroasiatic & vi & 21.2 & 24.1 \\
\bottomrule
\end{tabular}
\caption{
Model performance comparison with average MRR@10 across language families on mMARCO.
}
\label{tab:appendix}
\end{table}

\section{Effect of Alignment Granularity}
\label{F}

Table~\ref{tab:ablation_hsa} reports the effect of alignment granularity. Removing sentence-level and token-level alignment leads to MRR@10 drops of 1.2 and 0.7 points, respectively, showing that both granularities contribute and that sentence-level alignment plays a more dominant role. Nevertheless, token-level alignment remains valuable: by aligning keywords, entities, and local semantic units at a finer granularity, it compensates for the limitations of sentence-level alignment in fine-grained semantic modeling.

\section{Additional Analysis across Language Families and Resource Levels}
\label{G}
We use XLM-R$_{\mathrm{AGG}}$ as the main baseline and report absolute MRR@10 gains. For resource-level analysis, we exclude English and group the remaining 13 languages by the number of available EN--X parallel sentence pairs in WikiMatrix, whose statistics are shown in Figure~\ref{fig:f4}. The languages are sorted by this number and split into three roughly equal groups. We can observe that our method consistently outperforms XLM-R$_{\mathrm{AGG}}$ across all resource groups in Table~\ref{tab:resource_breakdown_mmarco}. Specifically, for high-resource languages, our method improves the average MRR@10 by 1.1 points, from 27.4 to 28.5. This indicates that even when sufficient parallel data is available, separating semantic and linguistic information can still further enhance cross-lingual semantic matching. Moreover, the medium-resource group obtains the largest gain of 1.9 points, suggesting that our method can further improve the shared semantic space when moderate parallel supervision is available. For low-resource languages, the stable gain of 1.2 points indicates that the model does not simply rely on large-scale parallel data, but can learn transferable language-agnostic representations through hierarchical semantic alignment and language-semantic disentanglement. 

To investigate model performance across different language families, we categorize the languages in mMARCO according to Glottolog's language-family classification. Table~\ref{tab:appendix} reports the results across different language families. Our method consistently outperforms XLM-R$_{\mathrm{AGG}}$ across all language groups, indicating that the proposed framework generalizes beyond a specific language family. Notably, clear gains are observed on typologically diverse languages such as Arabic, Chinese, Japanese, and Vietnamese, suggesting that disentangled semantic learning can reduce language-specific interference caused by differences in scripts, word order, and lexical forms. These results further support the effectiveness of separating language-agnostic semantic features from language-specific information for cross-lingual retrieval.

\section{Additional Experimental Results}
\label{H}
We present additional experimental results in this section.
For the main comparison with existing multilingual retrieval methods, Recall@1k results are reported in Table~\ref{tab:example-tabl_recall}, where our method still outperforms most baselines.
For out-of-domain evaluation, Recall@100 results are shown in Table~\ref{tab:example-tabl_recall100}. We observe consistently better performance from our method compared with other methods, demonstrating that our method generalizes well to out-of-distribution multilingual retrieval scenarios.

\begin{table*}[t]
\small
\renewcommand{\arraystretch}{1.26}
\centering

\resizebox{\textwidth}{!}{
\begin{tabular}{lccccccccccccccc}
\toprule
\textbf{Model} 
& \textbf{Avg} 
& \textbf{ar} 
& \textbf{de} 
& \textbf{en} 
& \textbf{es} 
& \textbf{fr} 
& \textbf{hi} 
& \textbf{id} 
& \textbf{it} 
& \textbf{ja} 
& \textbf{nl} 
& \textbf{pt} 
& \textbf{ru} 
& \textbf{vi} 
& \textbf{zh} \\
\midrule

\multicolumn{16}{l}{\textbf{Lexical Retrieval}} \\
\hdashline

BM25 
& 72.6 & 63.8 & 67.4 & 85.7 & 77.0 & 76.9 & 71.1 & 76.7 & 75.3 & 71.4 & 69.4 & 74.4 & 68.5 & 71.4 & 67.8 \\

\midrule

\multicolumn{16}{l}{\textbf{English Supervision}} \\
\hdashline

mDPR 
& 74.5 & 65.2 & 70.4 & 92.9 & 77.9 & 77.6 & 71.1 & 79.5 & 76.5 & 73.1 & 70.6 & 75.0 & 70.3 & 69.8 & 72.7 \\

mContriever 
& 76.6 & 70.1 & 80.5 & 92.1 & 77.3 & 77.2 & 72.4 & 79.6 & 76.8 & 72.7 & 76.6 & 77.4 & 76.2 & 71.5 & 71.9 \\

mColBERT 
& 77.1 & 70.9 & 83.3 & 92.8 & 77.5 & 76.2 & 75.4 & 76.8 & 79.1 & 74.4 & 76.5 & 77.8 & 76.3 & 71.6 & 71.3 \\

ColBERT-XM 
& 85.5 & 74.8 & 86.0 & \textbf{96.5} & 88.4 & 87.3 & 82.2 & 86.7 & 86.1 & 83.6 & 86.8 & 87.1 & 85.7 & 81.6 & \underline{84.8} \\

$\mathrm{mBERT}_{\mathrm{AGG}}$ 
& 83.9 & 72.9 & 83.6 & 94.1 & 87.9 & 86.7 & 81.2 & 85.8 & 86.3 & 80.5 & 84.4 & 86.2 & 84.5 & 78.0 & 82.5 \\

$\mathrm{XLM\mbox{-}R}_{\mathrm{AGG}}$ 
& 85.7 & 74.1 & 84.7 & 95.3 & 89.0 & 88.7 & 83.1 & 87.6 & 88.5 & 82.7 & 86.8 & \underline{87.5} & 86.1 & 80.2 & \underline{84.8} \\

\midrule

\multicolumn{16}{l}{\textbf{Multilingual Supervision}} \\
\hdashline

$\mathrm{mE5}_{\mathrm{BASE}}$
& 85.6 & 75.8 & 86.3 & 95.2 & 88.8 & \underline{90.1} & 82.2 & 86.4 & 87.6 & 83.8 & 86.3 & 86.0 & 84.1 & 83.2 & 82.4 \\

BGE
& 86.2 & 76.1 & 87.0 & 95.6 & 88.4 & \textbf{90.3} & 83.1 & 87.7 & 88.1 & 84.7 & \underline{87.4} & 86.8 & 84.9 & 83.8 & 83.3 \\

$\mathrm{BGE}_{\mathrm{HN}}$
& \underline{86.7} & \underline{76.6} & \underline{87.3} & \underline{96.0} & 88.9 & 89.7 & \underline{83.6} & \underline{88.5} & \textbf{88.9} & \textbf{85.1} & 87.2 & 87.3 & \underline{86.4} & \textbf{84.2} & 84.1 \\

\midrule

\multicolumn{16}{l}{\textbf{Ours}} \\
\hdashline

$\mathrm{Ours}_{\mathrm{mBERT}}$ 
& 85.8$^\dagger$ 
& 74.6$^\dagger$ 
& 86.4$^\dagger$ 
& 95.3$^\dagger$ 
& \underline{89.1}$^{\dagger\ddagger}$ 
& 88.5$^\dagger$ 
& 82.8$^\dagger$ 
& 87.4$^\dagger$ 
& 87.7$^\dagger$ 
& 84.0$^\dagger$ 
& 86.6$^\dagger$ 
& 86.5$^\dagger$ 
& 84.7$^\dagger$ 
& 82.1$^\dagger$ 
& \underline{84.8}$^\dagger$ \\

$\mathrm{Ours}_{\mathrm{XLM\mbox{-}R}}$ 
& \textbf{87.1}$^{\dagger\ddagger}$ 
& \textbf{76.9}$^{\dagger\ddagger}$ 
& \textbf{87.7}$^{\dagger\ddagger}$ 
& 95.8$^\dagger$ 
& \textbf{89.7}$^{\dagger\ddagger}$ 
& \underline{90.1}$^\dagger$ 
& \textbf{83.8}$^{\dagger\ddagger}$ 
& \textbf{89.3}$^{\dagger\ddagger}$ 
& \underline{88.7}$^\dagger$ 
& \underline{84.8}$^\dagger$ 
& \textbf{88.1}$^{\dagger\ddagger}$ 
& \textbf{87.9}$^{\dagger\ddagger}$ 
& \textbf{87.2}$^{\dagger\ddagger}$ 
& \underline{83.6}$^\dagger$ 
& \textbf{85.5}$^{\dagger\ddagger}$ \\

\bottomrule
\end{tabular}
}

\caption{In-domain zero-shot multilingual retrieval performance on the mMARCO dataset across 14 languages, measured by Recall@1k. 
$\dagger$ denotes significant improvements with paired t-test at $p<0.05$ over the baseline method using the same backbone model, and $\ddagger$ denotes significant improvements over the strongest baseline in the same column.
\textbf{Bold} and \underline{underline} indicate the best and second-best results, respectively.}
\label{tab:example-tabl_recall}
\end{table*}

\begin{table*}[t]
\centering
\renewcommand{\arraystretch}{1.25}
\setlength{\tabcolsep}{4.2pt}

\resizebox{\textwidth}{!}{
\begin{tabular}{l c cccccccccc c cccccccc c}
\toprule
\multirow{2}{*}{\textbf{Model}} 
& \multirow{2}{*}{\textbf{Avg}}
& \multicolumn{11}{c}{\textbf{With Parallel Data}}
& \multicolumn{9}{c}{\textbf{Without Parallel Data}} \\
\cmidrule(lr){3-13} \cmidrule(lr){14-22}
& 
& \textbf{ar} & \textbf{de} & \textbf{en} & \textbf{es} & \textbf{fr} 
& \textbf{hi} & \textbf{id} & \textbf{ja} & \textbf{ru} & \textbf{zh} 
& \textbf{Avg}$^{p}$
& \textbf{bn} & \textbf{fa} & \textbf{fi} & \textbf{ko} & \textbf{sw} 
& \textbf{te} & \textbf{th} & \textbf{yo} 
& \textbf{Avg}$^{np}$ \\
\midrule

\multicolumn{22}{l}{\textbf{Lexical Retrieval}} \\
\hdashline

BM25
& 67.3
& 78.7 & 42.8 & 63.6 & 25.4 & 50.2
& 73.8 & 71.8 & 73.6 & 56.4 & 55.1
& 59.1
& 90.0 & 68.1 & 81.2 & 70.1 & 69.9
& 73.3 & 87.5 & 80.1
& 77.5 \\

\midrule

\multicolumn{22}{l}{\textbf{English Supervision}} \\
\hdashline

mDPR
& 79.0
& 84.1 & 89.8 & 76.8 & \textbf{86.4} & \textbf{91.5}
& 77.6 & 57.3 & 82.5 & 79.7 & \textbf{94.4}
& 82.0
& 81.9 & \underline{89.8} & 78.8 & 73.7 & 61.6
& 76.2 & 67.8 & 71.5
& 75.2 \\

mContriever
& 84.9
& \textbf{92.5} & 84.1 & \underline{79.7} & 84.1 & 82.4
& 64.6 & 80.2 & 87.8 & 85.0 & 90.3
& 83.1
& 92.1 & 65.4 & \textbf{95.3} & 87.5 & \textbf{91.1}
& \textbf{96.1} & 93.6 & 77.0
& 87.3 \\

mColBERT
& 83.2
& \underline{90.8} & 80.3 & \textbf{80.1} & \underline{84.2} & 73.0
& \textbf{88.4} & 66.9 & \underline{89.5} & \textbf{86.6} & \underline{90.8}
& 83.1
& 91.3 & \textbf{91.0} & 83.2 & 72.2 & 69.2
& 83.0 & 84.5 & \textbf{91.7}
& 83.3 \\

ColBERT-XM
& 84.9
& 89.9 & 87.1 & 75.2 & 74.2 & 84.4
& 84.7 & 80.3 & 89.0 & 78.9 & 86.3
& 83.0
& 91.7 & 84.7 & 90.8 & 87.2 & 80.2
& 88.0 & 92.2 & 83.4
& 87.3 \\

$\mathrm{mBERT}_{\mathrm{AGG}}$
& 82.0
& 88.7 & 85.3 & 75.8 & 80.1 & 86.5
& 77.6 & 80.2 & 85.5 & 82.0 & 84.5
& 82.6
& 85.1 & 82.4 & 92.1 & 78.4 & 81.5
& 83.4 & 65.9 & 81.3
& 81.3 \\

$\mathrm{XLM\mbox{-}R}_{\mathrm{AGG}}$
& 86.0
& 90.2 & 88.4 & 75.4 & 78.2 & \underline{88.7}
& 82.5 & \textbf{87.6} & \textbf{89.8} & \underline{86.1} & 79.0
& 83.0
& \textbf{92.8} & 84.5 & \underline{93.1} & \underline{88.3} & 79.4
& \underline{95.5} & \textbf{94.5} & \underline{89.7}
& \textbf{89.7} \\






\midrule

\multicolumn{22}{l}{\textbf{Ours}} \\
\hdashline

$\mathrm{Ours}_{\mathrm{mBERT}}$
& \underline{86.4}$^\dagger$
& 90.3 & \underline{94.1}$^\dagger$ & 75.9 & 76.9 & 85.3
& 85.0 & 81.6 & 88.8 & 78.7 & 86.8
& \underline{84.3}$^\dagger$
& 92.4 & 85.2 & 92.3 & 88.0 & 82.5
& 92.9 & 93.5 & 84.2
& \underline{88.9} \\

$\mathrm{Ours}_{\mathrm{XLM\mbox{-}R}}$
& \textbf{87.3}$^\dagger$
& 90.7 & \textbf{94.8}$^\dagger$ & 76.5 & 79.6 & 86.0
& \underline{85.7} & \underline{82.5} & 89.4 & 79.3 & 88.9
& \textbf{85.3}$^\dagger$
& \textbf{92.8} & 85.8 & 93.0 & \textbf{88.7}$^\dagger$ & \underline{84.1}
& 93.8 & \underline{93.9} & 85.8
& \textbf{89.7} \\

\bottomrule
\end{tabular}
}

\caption{Out-of-domain zero-shot multilingual retrieval performance on the MIRACL dataset across 18 languages, measured by Recall@100. Languages are grouped according to whether their corresponding English–X parallel data are used during training. 
$\dagger$ denotes significant improvements with paired t-test at $p<0.05$ over the strongest baseline in the same column. 
\textbf{Bold} and \underline{underline} indicate the best and second-best results.}
\label{tab:example-tabl_recall100}
\end{table*}

\end{document}